\documentclass[aps,prb,reprint,superscriptaddress,showpacs]{revtex4-1}
\usepackage{graphicx}

\usepackage{amsmath}
\usepackage{amsfonts}
\usepackage{amssymb}
\usepackage{color}

\usepackage{bm}
\usepackage{sidecap}

\usepackage{epstopdf}
\usepackage{hyperref}

\begin{document}


\title{Photonic band gap of a graphene-embedded quarter-wave stack}
\author{Yuancheng Fan}
\email{phyfan@ameslab.gov, 429yuancheng@tongji.edu.cn}
\affiliation{Ames Laboratory and Department of Physics and Astronomy, Iowa State University, Ames, Iowa 50011, USA}
\affiliation{Key Laboratory of Advanced Micro-structure Materials (MOE) and \\ School of Physics Sciences and Engineering, Tongji University, Shanghai 200092, China}
\author{Zeyong Wei}
\affiliation{Key Laboratory of Advanced Micro-structure Materials (MOE) and \\ School of Physics Sciences and Engineering, Tongji University, Shanghai 200092, China}
\author{Hongqiang Li}
\email{hqlee@tongji.edu.cn}
\affiliation{Key Laboratory of Advanced Micro-structure Materials (MOE) and \\ School of Physics Sciences and Engineering, Tongji University, Shanghai 200092, China}
\author{Hong Chen}
\affiliation{Key Laboratory of Advanced Micro-structure Materials (MOE) and \\ School of Physics Sciences and Engineering, Tongji University, Shanghai 200092, China}
\author{Costas M. Soukoulis}
\email{soukoulis@ameslab.gov}
\affiliation{Ames Laboratory and Department of Physics and Astronomy, Iowa State University, Ames, Iowa 50011, USA}
\affiliation{Institute of Electronic Structure and Laser, FORTH, 71110 Heraklion, Crete, Greece}

\date{\today}

\begin{abstract}
Here, we present a mechanism for tailoring the photonic band structure of a quarter-wave stack without changing its physical periods by embedding conductive sheets. Graphene is utilized and studied as a realistic, two-dimensional conductive sheet. In a graphene-embedded quarter-wave stack, the synergic actions of Bragg scattering and graphene conductance contributions open photonic gaps at the center of the reduced Brillouin zone, that nonexistent in conventional quarter-wave stacks. Such photonic gaps show giant, loss-independent density of optical states at the fixed lower-gap-edges, of even-multiple characteristic frequency of the quarter-wave stack. The novel conductive sheets induced photonic gaps provide a new platform for the enhancement of light-matter interactions.

\end{abstract}

\maketitle

Photonic crystal is a type of artificial composite with photonic band gap (PBG) created by periodical modulated dielectric functions in the spatial domain. \cite{1,2,3,4,5} Conventional PBG stems from Bragg scatterings in the inhomogeneous media. By introducing different mechanisms into photonic crystals, it is shown that photonic band gaps can be modified, e.g., the zero-$\bar{n}$ gap, a consequence of stacking negative refractive index (negative-n) materials \cite{6} (artificial structures with simultaneously negative dielectric permittivities, $\epsilon$, and magnetic permeabilities, $\mu$). The PBG materials have found a myriad of applications to mold the flow of light, e.g., novel waveguides and photonic crystal fibers, microcavities, slow-light and photonic crystal lasers, directive emissions and collimation devices. \cite{7,8,9,10,11,12,13,14}

A monolayer carbon atoms in a hexagonal lattice with high-performance electric, thermal and mechanic properties \cite{15}---graphene, has recently attracted considerable attention for both its fundamental physics and enormous applications, e.g., in nanoelectronics. Such a realistic two-dimensional electron gas (2DEG) or two-dimensional (2D) material \cite{16,17} is also rising in photonics and optoelectronics. \cite{18} As a new optoelectronics material, graphene exhibits a much stronger binding of surface plasmon polaritons and supports its relatively longer propagation. \cite{19,20,21} Linear dispersion of the 2D Dirac fermions also provides ultrawideband tunability through electrostatic field, magnetic field, or chemical doping. \cite{18,22,23,24,25,26,27,28} However, since graphene is a monolayer atom sheet with  relatively low carrier concentrations, its interaction with light is not very strong, for example, graphene is almost transparent to optical waves. \cite{29} To strengthen graphene's ability for manipulating light signals, some artificially-constructed micro-structures have been introduced in a graphene sheet: graphene nanopatch has been exploited for complete optical absorption and nanospaser; \cite{30,31} dielectric super-lattices are utilized for the excitation of surface plasmon polaritons in graphene, \cite{32,33} and for hyperbolic metamaterials. \cite{34,35} In this paper, we propose a different mechanism for tailoring photonic band structures of a quarter-wave stack without changing its physical periods by embedding graphene sheets. The graphene conductance-induced photonic gap at the center of the reduced Brillouin zone, its fixed lower-gap-edge, and density of optical states that robust against losses in the graphene sheets are our primary interests.

The graphene-embedded quarter-wave stack for investigation is schematically illustrated in Fig. 1. A quarter-wave stack is composed of alternating high-index and low-index material layers with dielectric permittivities, $\epsilon_h$ and $\epsilon_l$ (with the thicknesses $d_h$ and $d_l$ ), which satisfies $\sqrt{\epsilon_h} d_h=\sqrt{\epsilon_l} d_l=\frac{\lambda_0}{4}$ ($\lambda_0$ is the wavelength in vacuum of characteristic frequency, $f_0$, of the quarter-wave stack. Here we set $f_0 = 2.5$ THz). The graphene sheets are embedded at interfaces of high-index and low-index layers, all dielectric layers and graphene sheets are lying on $xy$-planes. Note, throughout our study, the photonic crystals are illuminated by $x$-polarized (electric field, $E$, is along the $x$-axis) light with its propagation direction along the $z$-axis.

We employ a transfer matrix method (TMM) \cite{6,36,37,38} to investigate the optical properties of one-dimensional, layered structures embedded with conductive sheets. A recent study shows the one-dimensional layered system containing 2D materials such as monolayer graphene or 2DEG, can be solved by modifying the conventional transfer matrix method. That is accomplished by relating complex amplitude coefficients of forward and backward propagating electromagnetic fields in conductive sheets with the assistant of Ohm's Law. In the modified transfer matrix formalism, the propagation matrix in the dielectric layers is the same as conventional TMM: $M_p\left(\omega\right)=\text{diag}(e^{-i\phi_{h(l)}}, e^{i\phi_{h(l)}})$, $\phi_{h(l)}$ is the phase delay in each dielectric layer. Meanwhile, the transmission matrix changes to
\begin{equation}
\begin{split}
M_t\left(\omega\right)=\begin{bmatrix} 1+\eta_{h(l)}+\zeta_{h(l)} &1-\eta_{h(l)}-\zeta_{h(l)}\\1-\eta_{h(l)}+\zeta_{h(l)} &1+\eta_{h(l)}-\zeta_{h(l)}\end{bmatrix},
\end{split}
\end{equation}
with $\eta_{h(l)}=\frac{k_{z,l(h)}}{k_{z,h(l)}}$ and $\zeta_{h(l)}=\frac{\sigma_g \mu_{0} \omega}{k_{z,h(l)}}$. $k_{z,h(l)}$ is the $z$-component of the wave-vector, $k_{h(l)}=\sqrt{\epsilon_{h(l)}}\omega /c$ ($k=\omega /c$ is the wave-vector in vacuum), where $\omega$ is the angular frequency and $c$ is the speed of light in vacuum. $\zeta_{h(l)}$ is the contribution from the conductive sheet, which provides a new mechanism to tailor photonic band structure, as we will discuss below. Imposing a periodicity constraint, $E(z+a)=e^{i\kappa a}E(z)$ ($a$ is lattice constant of the the photonic crystal), leads to the solution condition for the complex Bloch wave-vector ($\kappa$) dispersion relation \cite{6,38}
\begin{equation}
2\cos(\kappa a)=\text{Tr}\left[T\left(\omega\right)\right]=\text{Tr}\left[\prod{M_p\left(\omega\right) M_t\left(\omega\right)}\right].
\end{equation}
\begin{figure}[t]
\includegraphics[width=8.6cm]{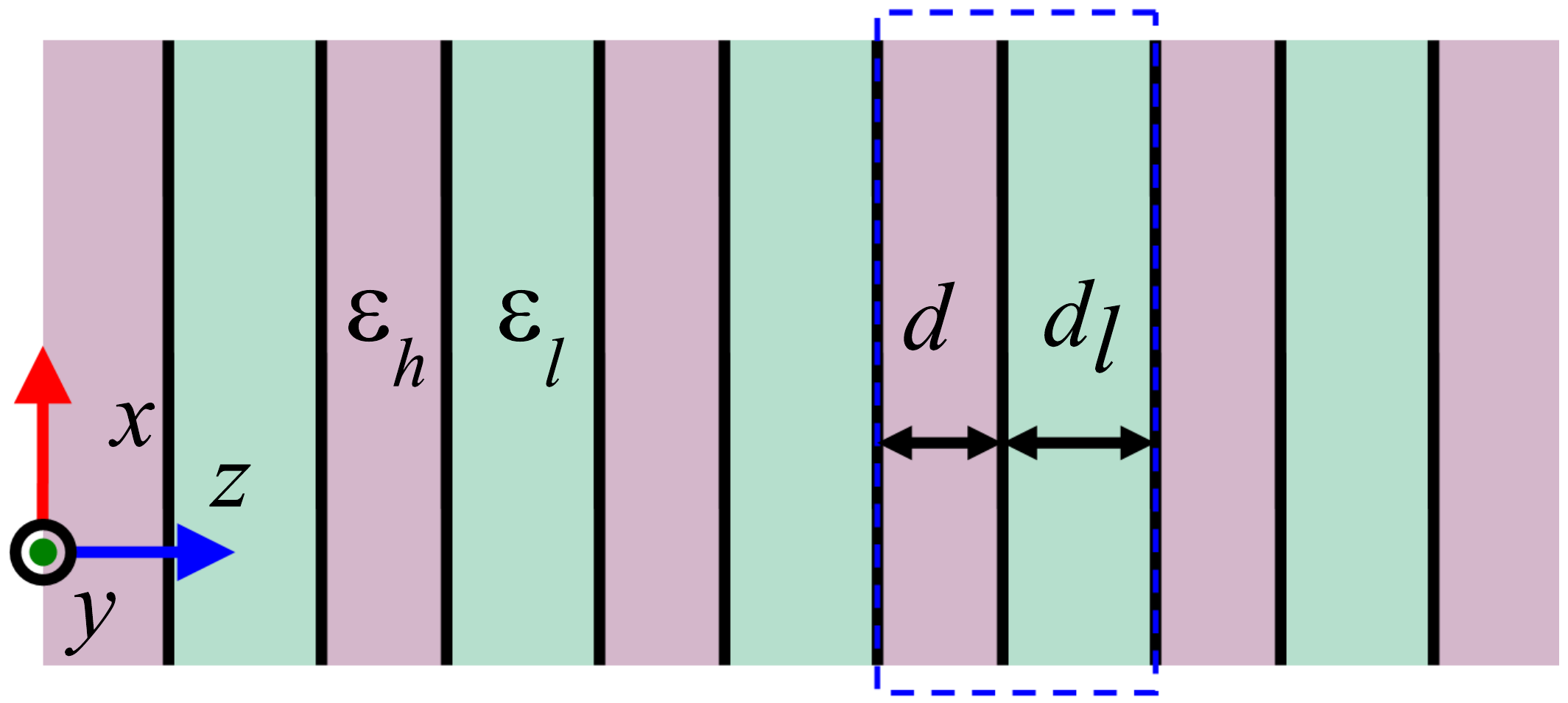}\caption{Schematic of the graphene-embedded photonic crystal in investigation: high-index ($\sqrt{\epsilon_h}$), low-index ($\sqrt{\epsilon_l}$) dielectric layers with thicknesses $d_h$ and $d_l$ are separated by graphene sheets (black lines), all are lying on the $xy$-plane, light propagates along $z$-direction.}\label{fg1}
\end{figure}

First, we consider a conventional quarter-wave stack with dielectric permittivities, $\epsilon_h = 2.9$ and $\epsilon_l = 1.5$.  The complex $\kappa$ dispersion relation is shown in the reduced zone scheme [Fig. 1(a)]. We can see photonic band gaps exist at the boundary ($\kappa = \pi/a$, or $k(\sqrt{\epsilon_h} d_h+\sqrt{\epsilon_l} d_l) = 1\pi, 3\pi, \cdots$) of the reduced Brillouin zone, while photonic band gaps does not exist at the center ($\kappa = 0$, or $k(\sqrt{\epsilon_h} d_h+\sqrt{\epsilon_l} d_l) = 2\pi, 4\pi, \cdots$ ) of the reduced Brillouin zone. \cite{4} Next, we take a realistic 2D material, the graphene as the conductivity sheet, into the quarter-wave stack, and study the band structure properties. The complex surface conductivity of the graphene sheet is adopted from a random-phase-approximation (RPA), \cite{39,40,41} which can be well described with a Drude model \cite{22,24} as $\sigma=i\alpha/\left(\omega+i\gamma\right)=i e^{2}E_{F}/\pi\hbar^{2}\left(\omega+i\tau^{-1}\right)$, especially at heavily doped region and low frequencies (far below Fermi energy), where $E_{F}=0.5$ eV is the Fermi energy. $\tau=\mu E_{F}/e\upsilon_{F}^2$ is the relaxation rate with $\mu=10^{4}$ cm$^2$V$^{-1}$s$^{-1}$, and $\upsilon_{F}\approx10^{6}$m/s the mobility and Fermi velocity, respectively. $E_{F}=\hbar \upsilon_{F}\sqrt{\pi\left|n\right|}$ can be easily controlled by electrostatic doping via tuning charge-carrier density $n$. In calculations, the theoretical graphene sheet was treated as an optical interface with complex surface conductivity ($\alpha = 58.86$ GHz/$\Omega$, $\gamma = \tau^{-1}= 2$ THz), since a one-atom-thick graphene sheet is sufficiently thin compared with the concerned wavelength. The complex $\kappa$ dispersion relation of a graphene-embedded quarter-wave stack is shown in Fig. 2(b). There exist photonic band gaps at the center of the reduced Brillouin zone, that is, the gaps start from 5 and 10THz. We also see the photonic gap begin with zero-frequency, due to the low frequencies, where the conductive graphene sheet blocks optical waves. We also note the gaps at the zone boundary are shifted toward high frequency compared with the quarter-wave stack without graphene. Furthermore, we found good agreement between the results for the theoretical model [in Fig.2(b) and Fig. 2(c)] and for graphene with the experimental data \cite{27,28} ($\alpha = 76.0$ GHz/$\Omega$, $\gamma = 9.8$ THz).

\begin{figure}[b]
\includegraphics[width=8.6cm]{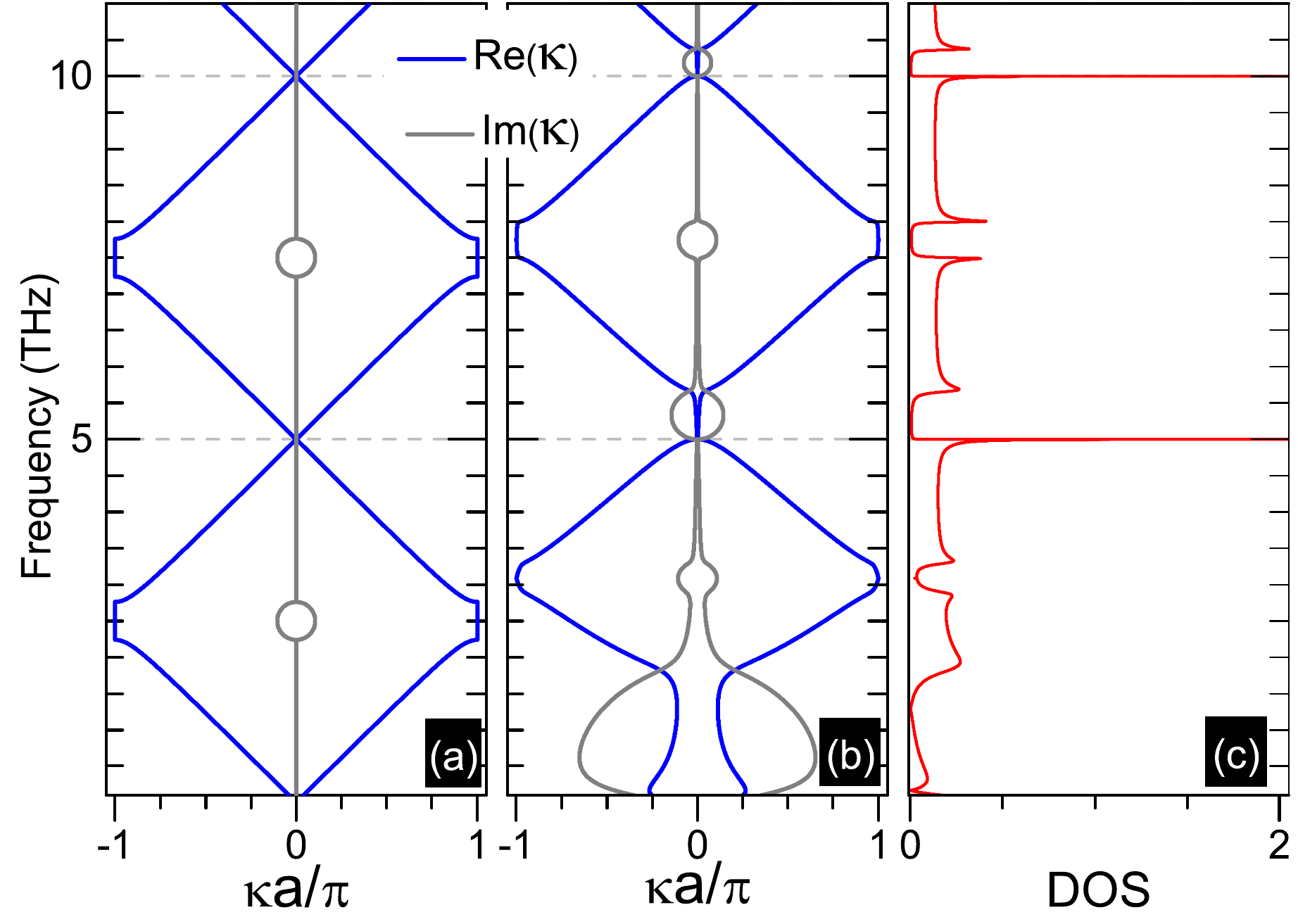}\caption{Comparison of the complex band structures of (a) a conventional quarter-wave stack,  and of (b) a graphene-embedded quarter-wave stack. (c) The density-of-optical-states (DOS) of the graphene-embedded quarter-wave stack corresponding to Fig. 2(b).}\label{fg2}
\end{figure}

For the graphene-embedded photonic crystal, we have the trace of the transfer matrix of the quarter-wave stack as:
\begin{equation}
\begin{split}
\text{Tr}\left[T\left(\omega\right)\right]=&\cos\left(\phi_h+\phi_l\right)\frac{1+\eta_h+\eta_l+\eta_h\eta_l}{2}\\
&+\cos\left(\phi_h-\phi_l\right)\frac{1-\eta_h-\eta_l+\eta_h\eta_l}{2}\\
&+\sin\phi_h \sin\phi_l\times\zeta_h\zeta_l\\
&-\sin\left(\phi_h+\phi_l\right)\frac{i\left(\zeta_h+\zeta_l+\eta_h\zeta_l+\eta_l\zeta_h\right)}{2}\\
&-\sin\left(\phi_h-\phi_l\right)\frac{i\left(\zeta_h-\zeta_l+\eta_h\zeta_l-\eta_l\zeta_h\right)}{2}.
\end{split}
\end{equation}
The first two terms are the contributions from Bragg scatterings in conventional photonic crystals. The last three terms contain permittivity associate impedance, $\eta$, and conductivity associate impedance, $\zeta$,  contributions, which represent synergic actions of Bragg scatterings and graphene surface conductance. That is the new mechanism introduced to modify the trace of the transfer matrix and, thus, the photonic band gaps. The trace spectra of the feature matrix for a conventional quarter-wave stack and a graphene-embedded quarter-wave stack are shown in Fig. 3 for comparison. In general, for a quarter-wave stack without graphene sheets (red curve), when
\begin{equation}
k(\sqrt{\epsilon_h} d_h+\sqrt{\epsilon_l} d_l) = m \pi, \quad  \left(m \in \textrm{integers}\right),
\end{equation}
the absolute value of the trace for the transfer matrix shows spectral maximum. To be more precise, if $m$ is an odd integer, then
\begin{equation}
\left|\text{Tr}\left[T\left(\omega\right)\right]\right| = \eta_{h}+\eta_{l}>2, \quad  \left(\epsilon_h \neq \epsilon_l\right).
\end{equation}
Equation (5) implies that Eq. (2) has no real solution for $\kappa$, indicating a spectral gap.
If $m$ is an even integer, then
\begin{equation}
\left|\text{Tr}\left[T\left(\omega\right)\right]\right| =1+ \eta_{h}\eta_{l}=2.
\end{equation}
Therefore, Eq. (6) implies that Eq. (2) has real solutions for $\kappa$. There is no spectral gap.
\begin{figure}[t]\includegraphics[width=8.6cm]{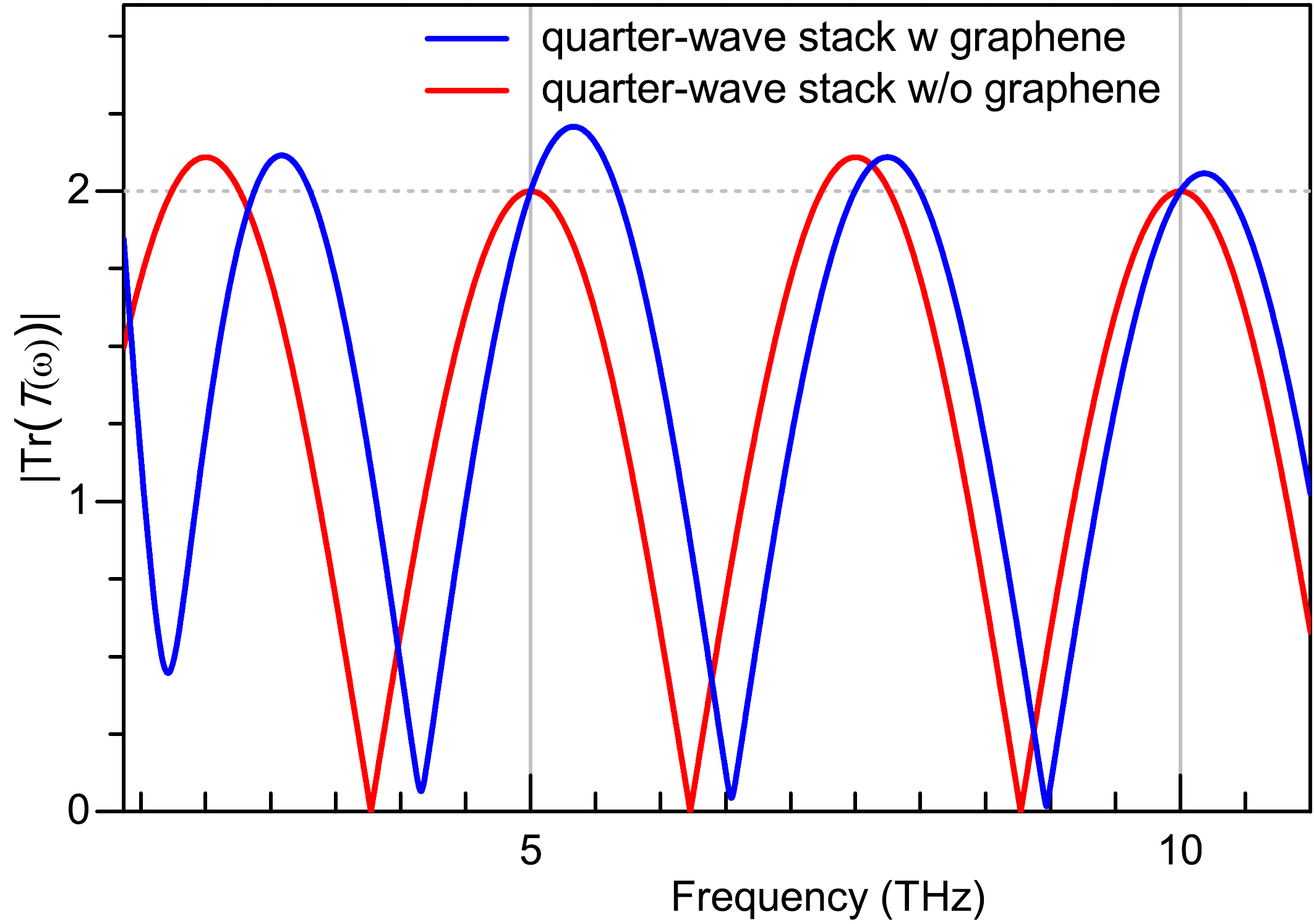}\caption{Spectra of the trace of the transfer matrices for a conventional quarter-wave stack and a graphene-embedded quarter-wave stack.}\label{fg3}
\end{figure}

Due to the joint contribution of graphene conductance and Bragg scatterings of the quarter-wave stack [last three terms on right side of Eq. (4)], there exist frequency ranges starting from $n f_0$ ($n \in \textrm{even integers}$), which imply new photonic band gaps--graphene sheet induced gaps at the center of the reduced Brillouin zone in Fig. 2(b). At frequency  $n f_0$, $\phi_h=\phi_l=\pi$, from Eq. (3) we obtain $\left|\text{Tr}\left[T\left(\omega\right)\right]\right|$ strictly equals $2$, which implies a fixed gap-edge. Starting with these fixed frequency points, there exist ranges where $\left|\text{Tr}\left[T\left(\omega\right)\right]\right|>2$ result from the contributions from graphene sheets, indicating the novel graphene induced gaps that are different from conventional photonic band gaps with their central frequencies fulfilling Bragg resonant conditions. The modifying of the graphene sheet on the gaps at $\kappa = \pi/a$ is also visualized in Fig. 3. We can see the blue-shift of the maximum in the spectra for the trace of the transfer matrices with the introduction of the graphene sheets. That corresponds to the blue-shift of the center of gaps at $\kappa = \pi/a$.

We know the light-matter interaction is determined to a great extent by the density-of-optical-states (DOS). The DOS (extracted from the band structure: $\frac{d \left|Re(\kappa)\right|}{d \omega}$) of graphene-embedded quarter-wave stack is shown correspondingly in Fig. 2(c). We can see there exist sharp peaks on the DOS spectrum at the fixed lower edge of the new graphene sheet induced gaps. Therefore we have reasons to believe the light-matter interaction at these frequencies will be significantly enhanced, compared to other frequencies, including the gap edge of other gaps. Usually, the DOS is sensitive to resistive loss in a photonic crystal. Here, in the graphene-embedded quarter-wave stack, loss exists in the conductive graphene sheets and, is represented by the collision frequency, $\gamma$. To elucidate the influence of the collision frequency of the graphene sheet on the DOS at resonance, Fig. 4 presents the calculated DOS spectra using a standard transfer matrix formalism for various collision frequencies (in unit of $\gamma_{0}$). As is shown in Fig. 4, there exist several peaks at the edges of the photonic gaps, when the collision frequency near $0$ for low-loss cases. Increasing the collision frequency of the graphene sheets, most of the peaks decay quickly. However, the DOS at the lower-edges of the new photonic gaps is unchanged with the increase of collision frequency.
\begin{figure}[b]\includegraphics[width=9.0cm
]{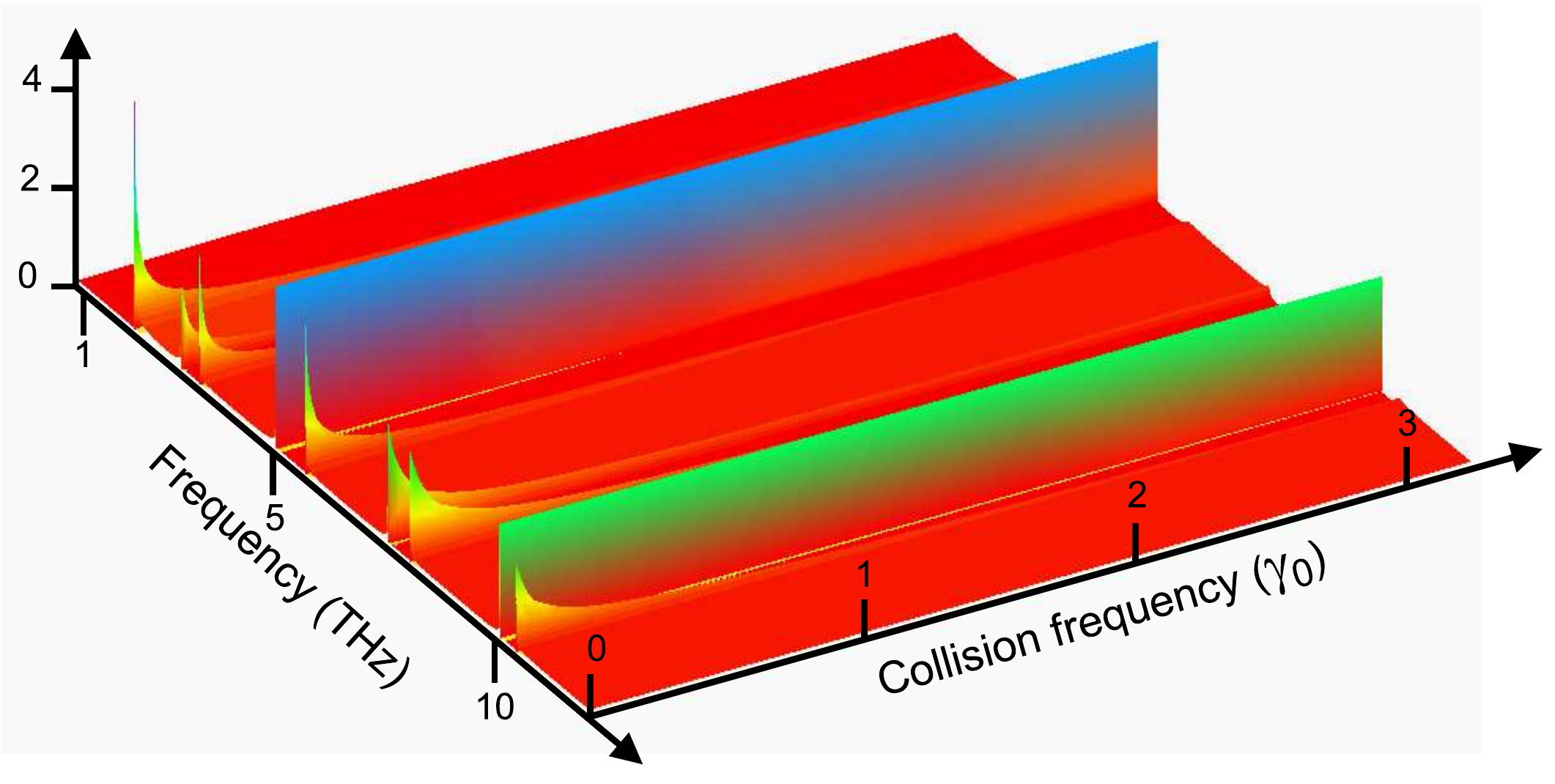}\caption{Density-of-optical-states (DOS) spectra of graphene-embedded quarter-wave stacks by changing the collision frequency (in unit of $\gamma_{0}$, $\gamma_{0} =2$ THz)  with other quantities of the graphene sheets as is stated above.}\label{fg3}
\end{figure}

To better understand the loss-independent nature of DOS at $n f_0$ ($n \in \textrm{even integers}$), the electric fields (amplitude) distribution of eigen modes at the lower-gap-edge was calculated and is depicted in Fig. 5. It is evident that standing waves exist in both high-index and low-index dielectric layers. The nodes of the standing wave locate at the interface of the high-index and low-index dielectric layers. As is shown in Fig. 5, the standing wave fields are with one loop and two loops inside each dielectric layer at the lower-gap-edge of the first (5.0 THz, upper panel) and the second (10.0 THz, lower panel) graphene-induced gap of the graphene-embedded quarter-wave stack. At these frequencies, we see electric fields showing localization within the two dielectric layers. Consequently, electromagnetic energy is efficiently suppressed at interface of  adjacent layers. Therefore, no fields locates in the graphene sheet, that gives rise to the pronounced loss-free DOS peaks in the collision frequency-dependent spectral map.

\begin{figure}[ptb]
\includegraphics[width=8.6cm]{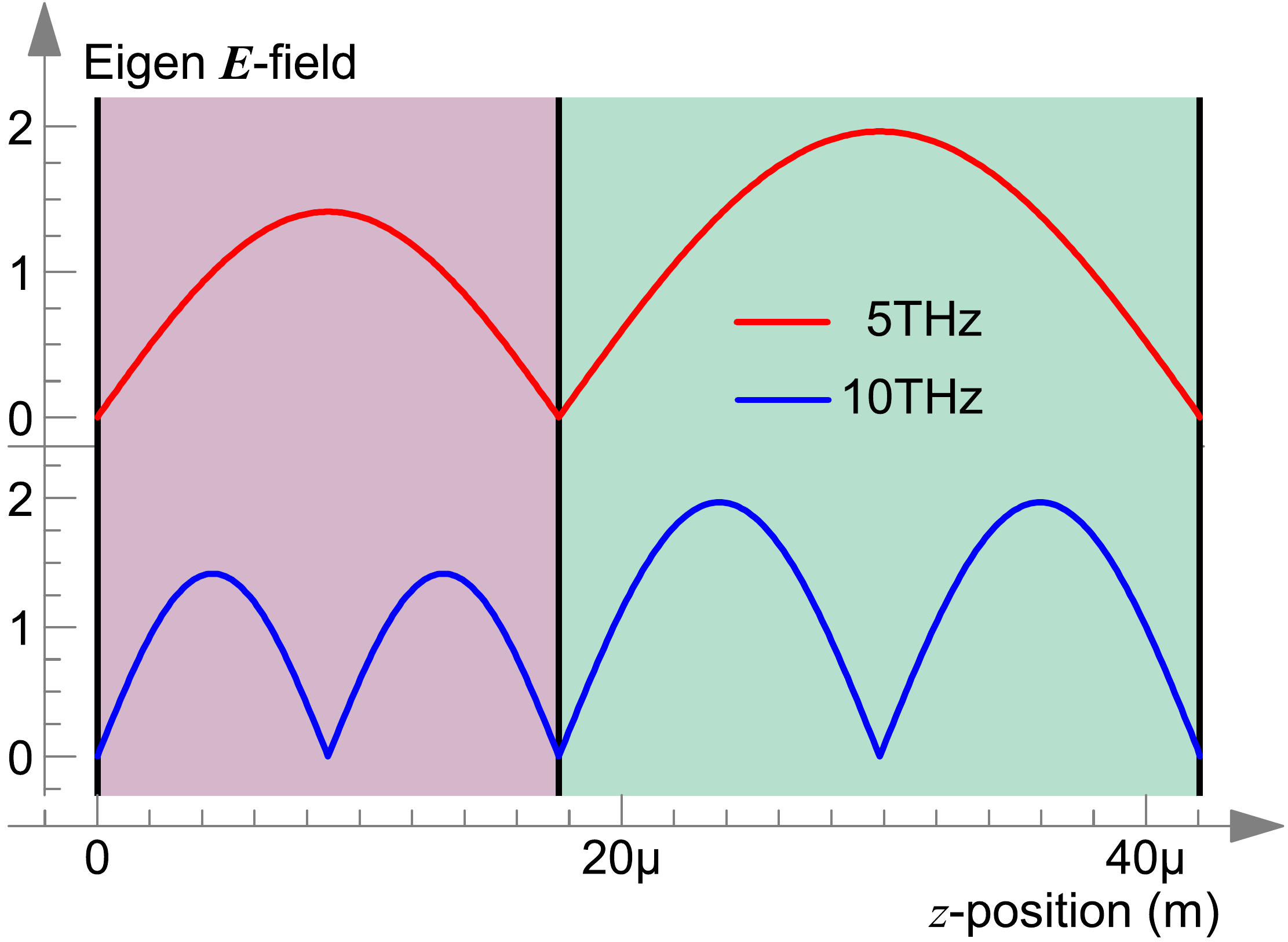}\caption{Electric fields (amplitude) of eigen modes at the lower-gap-edges of the first (5.0 THz, upper panel) and the second (10.0 THz, lower panel) graphene-induced gaps of the graphene-embedded quarter-wave stack are plotted for the unit cell (blue dashed box in Fig. 1). }\label{fg5}
\end{figure}

In essence, by introducing realistic conductive graphene sheets into quarter-wave stacks, we show the photonic band gap that does not exist at $\kappa = 0$ of the reduced Brillouin zone can now be opened with the assistance of the conductive graphene sheet with physical periods of the quarter-wave stack preserved, since the graphene sheet is optically thin. And, the new mechanism is not limited to graphene sheets. It is general for a 2D conductive material system category in the framework of 2DEG. That is a hot spot in recent condensed matter physics research. The research community has explored 2D materials beyond graphene, research of these 2D materials is rapidly expanding, with significant progress intensively reported from different material systems, such as nitrides (e.g., h-BN), dichalcogenides (e.g., MoS$_2$), topological insulators (e.g., Bi$_2$Se$_3$ or Bi$_2$Te$_3$), and even oxides. Moreover, density-of-optical-states at the lower-gap-edge is robust against loss of the conductive sheet. Such an advantage will benefit the development of platforms for enhancing light-matter interactions, which suffer from resistive losses. So far, the best platform for manipulating light-matter interaction have been photonic crystals composed of low-loss dielectric materials. Introduction of the conductive sheet discards this restriction, since the intrinsic losses in the 2D conductivity materials are insignificant to DOS. As a result, a conductive sheet can be introduced for modifying standing wave fields without accounting its inherent lossy nature in a photonic crystal scheme. The conductive sheet contributed giant, loss-independent DOS (or group delay), implies low group velocity, is beneficial for time-domain processing of optical signals, e.g., for the enhancement of nonlinear optical effects. \cite{10}

In conclusion, we suggest a novel mechanism for tailoring photonic band structures without changing physical periods of the photonic crystal structures. We have shown theoretically the photonic band gap in a quarter-wave stack opened with synergic actions of Bragg scattering and graphene conductance contributions. Its lower-gap-edges locate at even multiple characteristic frequency of the quarter-wave stack. Specifically, the giant DOS at the fixed lower-gap-edges is robust against losses in the graphene sheet, providing us with a new platform for potential applications to enhane light-matter interactions.

Work at Ames Laboratory was partially supported by the US Department of Energy (Basic Energy Sciences, Division of Materials Sciences and Engineering) under Contract No. DE-AC02-07CH11358. This work was partially supported by the Greek GSRT through the project ERC-02 EXEL. The work was supported by NSFC (Grants No. 11174221 and No. 10974144), CNKBRSF (Grant No. 2011CB922001). Y. Fan acknowledges Y. Chen and P. Zhang for helpful discussions and the China Scholarship Council (No. 201206260055) for financial support.

\end{document}